\documentstyle[aps,epsf,prd,eqsecnum]{revtex}

\begin{document}
\hspace{-10mm}
\vspace{-10.0mm} 
\thispagestyle{empty}
{\baselineskip-4pt
\font\yitp=cmmib10 scaled\magstep2
\font\elevenmib=cmmib10 scaled\magstep1  \skewchar\elevenmib='177
\leftline{\baselineskip20pt
\hspace{10mm} 
%
\rightline{\large\baselineskip20pt\rm\vbox to20pt{
\baselineskip14pt
\hbox{OU-TAP-144}
\vspace{1mm}
\hbox{\today}\vss}}%
}
\vskip8mm
\begin{center}{\large \bf
Brane-world inflation without inflaton on the brane} 
\end{center}
\vspace*{4mm}
\centerline{\large 
Yoshiaki Himemoto\footnote{E-mail:himemoto@vega.ess.sci.osaka-u.ac.jp}
and Misao Sasaki\footnote{E-mail:misao@vega.ess.sci.osaka-u.ac.jp}}
\vspace*{4mm}
\centerline{\em Department of Earth and Space Science, Graduate 
School of Science}
\centerline{\em Osaka University, Toyonaka 560-0043, Japan} 

\begin{abstract}
Inspired by the Randall-Sundrum brane-world scenario, we investigate
the possibility of brane-world inflation driven not by
an inflaton field on the brane, but by a bulk, dilaton-like gravitational
field. As a toy model for the dilaton-like gravitational field,
we consider a minimally coupled massive scalar field in the bulk
5-dimensional spacetime, and look for a perturbative solution in the
anti-de Sitter (AdS) background. For an adequate range of the scalar field 
mass, we find a unique solution that has non-trivial dependence on the
5th dimensional coordinate and that induces slow-roll inflation on the
brane.
\end{abstract}
\vskip10mm

\section{Introduction}
It is now widely accepted that our spacetime is not
4-dimensional but higher dimensional
from the unified theoretical point of view.
Horava and Witten showed the possibility that desirable 
gauge fields may consistently appear on the 10-dimensional boundary of
$Z_2$-symmetric 11-dimensional spacetime \cite{Horava}. 
A low-energy, 5-dimensional realization of the Horava-Witten theory was
first discussed by Lukas et al. \cite{lukas}.
They also analyzed brane-world cosmology in the
context of this 5-dimensional theory \cite{lukascos1,lukascos2}.

Recently, Randall and Sundrum wrote two very interesting papers
\cite{rs1,rs2} on another possible low-energy realization of the
Horava-Witten theory.
In \cite{rs1}, they found an interesting $Z_2$-symmetric
solution of the 5-dimensional Einstein equations with a 
negative cosmological constant.
In this solution, two boundary branes with positive and negative
tensions are embedded in the 5-dimensional
anti-de Sitter (AdS) space and the tensions of the branes are chosen
so that the effective cosmological constant on the branes vanishes
and the 4-dimensional Minkowski space is realized on the branes.
They then showed that the mass-hierarchy problem in particle physics
may be solved if we live on the negative tension brane.
This work has received much attention from the particle physics
community, and subsequently a large number of papers have been published
on it. However, it was soon realized that there exists the so-called
radion mode that describes fluctuations of the distance between the two
branes, and this mode causes an unacceptable modification of the
effective gravitational theory on the negative tension brane,
unless the radion is stabilized rather artificially\cite{tanaka,tX,rubakov}.

In the second paper\cite{rs2}, Randall and Sundrum showed that
the negative tension brane may actually be absent if we live on
the positive tension brane. They found because of the curvature of AdS,
even if the extra-dimension is infinite, gravity on the brane
is nicely confined around the brane and the Einstein gravity is
effectively recovered on the brane\cite{rs2,tanaka,sms,gid}.
As a result, although the hierarchy problem remains unsolved, this
model has attracted much attention from the relativity/cosmology
community, and the brane-world cosmology has
boomed\cite{flanagan,pierre,kraus,muko,ida,%
stoica,kodama,david,bruck,gong,yu,koyama,kim,dorca,sasaki,luis,olsen,roy}.

To name a few, the Friedmann equation on the brane
is discussed in \cite{kraus,muko,ida,kim,dorca,sasaki,roy}.
The quantum creation of a brane-world is
discussed in \cite{sasaki,luis}.
The formulation and evolution
of cosmological perturbations in the brane-world are discussed
in \cite{kodama,david,bruck,koyama,dorca}. 
And inflation on the brane is discussed
by many people \cite{sasaki,luis,olsen,kazuya,a,b,c,d,e,f,g,h}. 

In almost all of these works, the 5-dimensional bulk spacetime 
is assumed to be vacuum except for the presence of the cosmological
constant, and the matter fields on the brane are regarded as
responsible for the dynamics of the brane.
 However, from the unified theoretic point of view, 
the gravitational action is not necessarily the Einstein-Hilbert
action. In fact, string theory tells us that 
the dimensionally reduced effective action includes not only
higher-order curvature terms but also dilatonic gravitational scalar
fields. Thus at the level of the ``low-energy'' 5-dimensional theory,
it is naturally expected that there appears a dilaton-like scalar field
in addition to the Einstein-Hilbert action \cite{lukas}.
Thus it is of interest to investigate how such a scalar field in
the 5-dimensional theory affects the brane-world.
In this connection, very recently Maeda and Wands have investigated 
dilaton-gravity in the brane-world scenario \cite{wands}.

In this paper, we investigate whether it is possible or not to have
inflation on the brane solely by a bulk gravitational scalar field.
Thus we consider a scenario of inflation
on the brane without introducing an inflaton field on the 
brane.\footnote{This type of inflation is called ``modular inflation''
in \cite{lukascos2}.}
A similar model of inflation in 5-dimensional dilatonic gravity
was discussed recently by
Nojiri, Obregon and Odintsov\cite{NoObOd}.
We model the dilaton-like bulk gravitational scalar by
a minimally coupled scalar field. It is known that 
a scalar-tensor gravitational theory is conformally 
equivalent to the Einstein theory plus a minimally coupled scalar
field\cite{maeda}.
So our model may be regarded as a conformally transformed 
scalar-tensor gravitational theory.

Very recently, the quantum fluctuations in a similar brane-world
scenario has been discussed by Kobayashi, Koyama and Soda\cite{shinpei}.
They have shown that the extra-dimensional corrections due to
the so-called Kaluza-Klein modes are small and the scenario is
observationally acceptable just as the ordinary 4-dimensional
scenario of inflation.

This paper is organized as follows. In Sec.~II, we first recapitulate
the effective 4-dimensional gravitational equations on the brane
and derive the Friedmann equation on the brane under the presence
of a bulk scalar field. Then we discuss general conditions for inflation
to occur on the brane by the bulk scalar field.
In Sec.~III, we solve the Einstein-scalar field equations in the bulk
perturbatively, and give explicit solutions that give rise to
inflation on the brane.
Finally, we summarize our results and discuss the implications
in Sec.~IV.

\section{Effective 4-dimensional equations on the brane}\label{4Deq}
In order to understand the general conditions required to induce
inflation on the brane, we review the effective 4-dimensional
gravitational equations, and derive the Friedmann equation under the
presence of a bulk scalar field. 

\subsection{The Einstein equations on the brane}
The general form of the effective gravitational equations on the brane
was discussed in \cite{sms}. Following it, we consider the 
metric near the brane in the Gaussian normal coordinates,
\begin{equation}
ds^{2}  = g_{a b} dx^{a}dx^{b}=d\chi^{2}+q_{\mu \nu}dx^{\mu}dx^{\nu}\,,
\label{gmetric}
\end{equation}
where the brane is located at $\chi=\mbox{constant}$. Without loss of
generality, we set this constant at zero.
We assume that the 5-dimensional bulk gravitational equation
takes the form,
\begin{equation}
  \label{bulkgreq}
  R_{ab}-{1\over2}g_{ab}R+\Lambda_5g_{ab}
=\kappa_5^2\left(T_{ab}+S_{a b} \delta(\chi)\right)\,,
\end{equation}
where the 5-dimensional energy-momentum tensor $T_{ab}$ 
stands for a dilaton-like scalar field.
For our toy model, we assume it is given by a minimally coupled
scalar field,
\begin{equation}
T_{a b} = \phi_{,a} \phi_{,b}- g_{a b}\left( {1\over2} 
g^{c d}\phi_{,c} \phi_{,d}+ V(\phi)\right).
\label{5emt}
\end{equation}
The energy-momentum tensor on the brane $S_{ab}$
is assumed to be of the form,
\begin{equation}
S_{a b} = -\sigma q_{a b}+\tau_{a b} \,;
\quad \tau_{ab}n^b=0\,,
\end{equation}
where $n^a$ is the vector unit normal to the brane,
$\sigma$ is the brane tension and 
$\tau_{ab}$ describes the 4-dimensional matter fields.
The field equation for the bulk scalar $\phi$ is
\begin{equation}
\Box_{(5)}\phi -V'(\phi)=
\partial_{\chi}^{2}\phi + \Box_{(4)}\phi-V'(\phi)=0.
\label{5fe}
\end{equation}

According to \cite{sms}, the induced 4-dimensional Einstein tensor 
${}^{(4)}G_{ab}$ on the $\chi=\mbox{constant}\neq0$ hypersurface 
is given by
\begin{eqnarray}
{}^{(4)}G_{a b} & = & {2 \kappa_5^{2}\over3}
\left[(T_{cd}-{\Lambda_5 \over{\kappa_5^{2}}}g_{c d}) 
q_a{}^c q_b{}^d+
\left((T_{cd}- {\Lambda_5 \over{\kappa_5^{2}}}g_{c d})
n^{c} n^{d}
-{1\over 4}(T^c{}_c
-{5\Lambda_5 \over{\kappa_5^{2}}})\right) q_{ab}\right] \nonumber \\
& &-E_{ab}
+K K_{ab}-K_{a}{}^{c}K_{bc}-{1\over2}q_{ab}(K^2-K^{cd}K_{cd}),
\label{4ein}
\end{eqnarray}
where $K_{ab}=q_a{}^c n_{c;d}q^d{}_b$ is the extrinsic curvature of the
$\chi=\mbox{constant}$ hypersurface, and
$E_{a b}=C_{acbd}n^c n^d$ where $C_{acbd}$ is the
5-dimensional Weyl tensor.

Following the spirit of the RS brane-world, we assume
the bulk spacetime is $Z_2$ (reflection) symmetric with respect
to the brane. 
This implies the junction condition,
\begin{equation}
[K_{\mu \nu}]=2K_{\mu\nu}^{+}=-2K_{\mu\nu}^{-}
=-\kappa_5^{2}\left({\sigma \over 3} q_{\mu\nu}+\tau_{\mu\nu}
-{1\over3}q_{\mu\nu}\tau\right),
\end{equation}
where $[K_{\mu \nu}]=K_{\mu \nu}^{+}-K_{\mu \nu}^{-}$.
Substituting Eq.(\ref{5emt}) into Eq.(\ref{4ein}), 
and taking the limit $\chi\rightarrow\pm 0$,
${}^{(4)}G_{a b}$ on the brane becomes
\begin{equation}
G_{\mu \nu}=-\Lambda_{4}q_{\mu \nu} 
+\kappa_{4}^{2}(T_{\mu \nu}^{(s)}+\tau_{\mu \nu})+
\kappa_{5}^{4} \pi_{\mu \nu}-E_{\mu \nu},
\end{equation}
where
\begin{eqnarray}
\Lambda_{4} & = & {1\over2} 
\left(\Lambda_5+{1\over6}\kappa_5^{4}\sigma^{2}\right),\\
\kappa_{4}^{2}&=&{\kappa_{5}^{4}\sigma \over 6}, \\
T_{\mu \nu}^{(s)} & = &{1\over{\kappa_5^{2} \sigma}}
\left(4 \phi_{,\mu} \phi_{,\nu}+\left({3\over2}(\phi_{,\chi})^{2}
-{5\over2}q^{\alpha \beta}\phi_{,\alpha} \phi_{,\beta} -3V(\phi)\right)
q_{\mu \nu}\right),\\
\pi_{\mu \nu} &=& -{1\over4}\tau_{\mu \alpha}\tau_{\nu}^{\ \alpha}
+{1\over12}\tau \tau_{\mu \nu} 
+{1\over8}q_{\mu \nu}\tau_{\alpha \beta}\tau^{\alpha\beta}
-{1\over24}q_{\mu \nu}\tau^{2}\,.
\end{eqnarray} 

\subsection{The Friedmann equation on the brane}
We consider a spatially isotropic, homogeneous universe on the brane:
\begin{equation}
\left.ds^{2}\right|_{\chi=0}
=q_{\mu\nu}(\chi=0)dx^\mu dx^\nu
=-dt^{2}+a(t)^{2}\gamma_{i j}dx^{i} dx^{j},
\end{equation}
where $\gamma_{ij}$ is the metric of a constant curvature space
with curvature $K=\pm1$, $0$.
We assume the matter fields on the brane satisfy the energy-momentum
conservation law $D_\nu \tau^{\mu\nu}=0$, 
where $D_{\nu}$ is the covariant derivative with respect to
$q_{\mu\nu}$.
This implies\cite{sms} 
\begin{eqnarray}
D_{\nu} \tau_{\mu}{}^{\nu} 
\propto D_{\nu}K_{\mu}{}^{\nu}-D_{\mu}K &=& \kappa_5^{2} 
{T}_{ab}n^{a} q^b{}_{\mu}
= \kappa_5^{2} T_{\chi\mu} 
= 0\,.
\end{eqnarray}
Hence, the bulk scalar $\phi$ is even with respect to $\chi$,
i.e., $\phi_{,\chi}|_{\chi=0}=0$.

We assume the perfect fluid form for $\tau_{\mu\nu}$
on the brane:
\begin{equation}
\tau^{\mu \nu}=\rho^{(m)}t^{\mu}t^{\nu}+Ph^{\mu \nu}.
\end{equation}
Then, the Friedmann equation on the brane becomes
\begin{equation}
3\left[\left({\dot a\over a}\right)^{2}+{K\over a^2}\right]
=\kappa_4^{2}(\rho^{(s)}+\rho^{(m)})+
{\kappa_{5}^{4} \over12}\rho^{(m)2}-E_{tt}+\Lambda_{4},
\label{friedmann}
\end{equation}
where
\begin{equation}
\rho^{(s)}={3\over{\kappa_5^{2}\sigma}}\left({\dot{\phi}^{2}\over2}
+V(\phi)\right)={1\over2}\dot{\Phi}^{2}+\tilde{V}(\Phi).
\end{equation}
Here, we have introduced the effective scalar field $\Phi$ 
on the brane by rescaling the bulk scalar field as
\begin{equation}
\Phi:=\sqrt{3\over{\kappa_5^{2} \sigma}}\phi.
\end{equation}

Now we examine if there is a situation in which all but the $\rho^{(s)}$
term on the right-hand-side of Eq.~(\ref{friedmann}) can be neglected,
and inflation occurs due to the potential $\tilde V(\Phi)$. For
simplicity, we neglect the matter terms;
their role is the same as the standard 4-dimensional theory
except for the presence of the term quadratic in $\rho^{(m)}$.
As for the $\Lambda_4$ term, we assume it is fine-tuned to a very small
value or it can be just absorbed into the $\tilde V(\Phi)$ term.
In any case, the cosmological constant problem is beyond the scope of
this paper.
Then, the remaining, possibly dangerous term is the $E_{tt}$ term.
Since it carries information of the bulk gravitational field,
it cannot be determined solely by the 4-dimensional equations.
Nevertheless, it is possible to obtain some general features of the
term by considering the Bianchi identities.

As discussed in \cite{sms}, the spatial homogeneity of the brane
implies
\begin{equation}
D_{\mu}\pi^{\mu \nu}=0.
\end{equation}  
Then, the 4-dimensional Bianchi identities imply
\begin{equation}
\kappa_4^{2} D^{\mu} T_{\mu \nu}^{(s)}=D^{\mu} E_{\mu \nu}.
\label{hozon}
\end{equation}
Note that $E_{\mu \nu}$ is traceless i.e. $E_t^{\,t}=-E_i^{\,i}$.
The only non-trivial component of the above equation in the present
case is the time-component, which becomes 
\begin{equation}
{\kappa_5^{2}\over6}
(\ddot{ \phi}-4\,\partial_{\chi}^{2}\,\phi+V') \dot{\phi}
=-{1\over a^{4}}\partial_{t}(a^{4} E_{tt}).
\end{equation}
Using the 5-dimensional field equation (\ref{5fe}), this is
rewritten as
\begin{equation}
{\kappa_{5}^{2}\over2}(\partial_{\chi}^{2}\phi 
+ {\dot a\over a} \dot{\phi})\dot{\phi}=
{1\over{a^{4}}}\partial_{t}(a^{4} E_{tt}).
\end{equation}
Therefore 
\begin{equation}
E_{tt}={\kappa_5^{2}\over{2a^{4}}}\int^t a^{4}\dot{\phi}
(\partial_{\chi}^{2}\phi + {\dot a\over a}\dot{\phi})\,dt.
\label{weyl}
\end{equation}
The integration constant in the above integral gives rise to
the `dark radiation' term proportional to $a^{-4}$.
Hence we may neglect it if inflation should occur.
We then see that $E_{tt}$ can be neglected if both $\dot\phi$ and
$\partial_\chi^2\phi$ are sufficiently small, i.e.,
\begin{eqnarray}
  \label{phicond}
  \dot\phi^2\ll V(\phi) \,,\quad
 |\partial_\chi^2\phi|\leq {\dot a\over a}|\dot\phi|\,,
\end{eqnarray}
on the brane.

Thus a sufficient condition for inflation to occur on the brane
is that $\phi$ is a slowly varying function with respect to
both $t$ and $\chi$ in the vicinity of the brane.
In the next section, we look for a solution with such a
property.

\section{The solution in the bulk}\label{bulksol}
We want to find a solution of the field equations that 
has non-trivial dynamics in the bulk and gives rise to
inflation on the brane. Since such a solution will naturally
have non-trivial dependence on $t$ and $\chi$, we take a perturbative
approach.

\subsection{Model}

It has been pointed out in \cite{sasaki} that
the metric for a natural brane universe model in the cosmological
context takes the form,
\begin{equation}
ds^{2}=dr^{2}+(H\ell)^{2} \sinh^{2}(r/\ell)
[-dt^{2}+H^{-2}\cosh^{2}(Ht)d\Omega_{(3)}^{2}] \quad  (r \leq r_{0}),
\label{metric}
\end{equation}
where $d\Omega_{(3)}^{2}$ is the metric on the unit 3-sphere,
$r_0$ is the location of the brane
and the brane asymptotically inflates with the Hubble rate $H$,
\begin{equation}
H(\ell)={1\over\ell\sinh (r_0 /\ell)}\,;
\quad \ell=\left|{6\over{\Lambda_5}}\right|^{1/2}.
\end{equation}
The brane tension $\sigma$ is given by
\begin{equation}
  \sigma=\sigma_c\coth(r_0/\ell)\,;
\quad \sigma_c={6\over{\kappa_5^{2} \ell}}\,,
\end{equation}
where $\sigma_c$ is the critical tension that gives $\Lambda_4=0$
and reproduces the original RS brane-world\cite{rs2}.
A nice feature of this model is that it can be interpreted as
the brane universe created from nothing, and the universe naturally
inflates after creation.

Here we essentially adopt this model, but slightly modify
the interpretation of the model parameters and 
consider the following scenario.
We assume that the 5-d cosmological constant $\Lambda_5$ 
and the brane tension $\sigma$ are determined from some yet unknown
unified theory such that $\sigma=\sigma_c$.
However, due to dynamics of the bulk gravitational field, the
dilaton-like scalar field with an effective
potential $V(\phi)$ appears in the bulk. 
We assume $V>0$ which may vary very slowly in space and time.
The brane-world is created from nothing in this situation.  Then, as
will be discussed in detail below, inflation occurs on the brane.
As the universe evolves, the potential $V$ decreases
and eventually becomes zero (or very small) and the standard Friedmann
universe is recovered in the low energy limit.

The specific model we consider is as follows.
For simplicity, we assume the potential of the form,
\begin{equation}
V(\phi)=V_0+{1\over2}m^{2}\phi^{2},
\end{equation}
with $m^{2}<0$ and consider the situation when $\phi$ is
sufficiently close to zero.
In the lowest order, we put $\phi=0$.
Then the 5-dimensional bulk spacetime is AdS with
the cosmological constant $\Lambda_{5,eff}$,
where
\begin{equation}
\Lambda_{5,eff}=\Lambda_5+\kappa_5^{2}V_0 \,.
\label{Lambda5eff}
\end{equation}
The Friedmann equation on the brane becomes
\begin{equation}
\left({\dot a\over a}\right)^2
+{K\over a^2}=H^{2}
={\kappa_5^{2}V_0\over6}\,,
\label{walker}
\end{equation}
where $K=1$. Note that $V_0$ must satisfy the condition 
$|\Lambda_5|>\kappa_5^{2}V_0$ so that the background remains 
effectively AdS, i.e., $\Lambda_{5,eff}<0$.
Thus the lowest order bulk spacetime has the metric
(\ref{metric}) with $\ell$ replaced by $\ell_{eff}$,
where
\begin{equation}
\ell_{eff}^{2}={6\over\left|\Lambda_{5,eff}\right|}.
\label{effective}
\end{equation}

On this effective AdS background, we look for a perturbative
solution for $\phi$. Assuming the solution is spherically
symmetric, the field equation in the bulk becomes
\begin{equation}
{1\over{H^{2}l_{eff}^{2}\sinh^{2}(r/l_{eff})\cosh^{3}(Ht)}}
\partial_{t}\left(\cosh^{3}(Ht)\partial_{t}\phi\right)
-{1\over{\sinh^{4}(r/l_{eff})}}\partial_{r}\left(\sinh^{4}(r/l_{eff})
\partial_{r}\phi\right)+m^{2}\phi=0\,,
\label{5feq}
\end{equation}
with the boundary condition $\partial_r\phi=0$ at $r=r_0$.
Once the solution is found, we can solve the Friedmann equation
(\ref{friedmann}) together with the equation (\ref{weyl}) for $E_{tt}$
with the identification $\chi=r-r_0$,
to find the cosmological evolution on the brane perturbatively. 

\subsection{Slow-roll condition}
Our task is to find a regular solution of Eq.~(\ref{5feq})
that satisfies the slow-roll condition on the brane.
{}From Eq.(\ref{effective}), 
\begin{equation}
H^{2}\ell_{eff}^{2} 
= {1\over{\sinh^{2}(r_0/l_{eff})}}= {1\over{z_0^{2}-1}},
\end{equation}
where, $z_0 =\cosh (r_0/l_{eff})$. Using Eqs.~(\ref{Lambda5eff}) and
(\ref{walker}), the above equation can be expressed as
\begin{equation}
H^{2}\ell_{eff}^{2} 
={\kappa_5^{2} V_0\over{|\Lambda_5|-\kappa_5^{2} V_0}}=
{1\over{|\Lambda_5|/\kappa_{5}^{2}V_0}-1}. 
\end{equation}
Hence the position of brane is determined by the ratio between the
5-dimensional cosmological constant $\Lambda_5$ and the vacuum energy
$V_0$,
\begin{equation}
z_0^{2}=\cosh^2(r_0/\ell_{eff})
={|\Lambda_5|\over \kappa_5^{2} V_0}>1.
\end{equation}
Note that the last inequality is the condition for the bulk spacetime 
to be AdS. 
{}From above the relation, we find
\begin{equation}
{|m^{2}|\over H^2}=|m^{2}|l_{eff}^{2}(z_0^{2}-1).
\label{srcondition}
\end{equation}
We expect slow-roll inflation to occur if $|m^2|/H^2\ll 1$.

\subsection{Solution}\label{solution}
To solve the field equation (\ref{5feq}),
we look for a solution of the separable form,
\begin{equation}
\phi(t,r)=\psi(t)\,u(r)\,.
\end{equation}
Then we obtain
\begin{equation}
\left[-{1\over{\sinh^{4}(r/l_{eff})}}\partial_{r}\sinh^{4}(r/l_{eff})
\partial_{r} 
+m^{2}+{\lambda^{2}\over{l_{eff}^{2}}
\sinh^{2}(r/l_{eff})}\right] u(r)=0,
\label{req}
\end{equation}
\begin{equation}
\left[{1\over{\cosh^{3}(Ht)}}\partial_{t}\cosh^{3}(Ht)\partial_{t}
-H^{2}\lambda^{2} \right]\psi(t)=0,
\label{timeeq}
\end{equation}
where $\lambda^{2}$ is the separation constant.
The general solutions of Eqs.~(\ref{req}) and (\ref{timeeq}) are
given by
\begin{eqnarray}
u=u_{\gamma}(r) &=& 
A{P_{\nu-1/2}^{-\gamma-3/2}(\cosh (r/\ell_{eff}))
\over{\sinh^{3/2}(r/\ell_{eff})}} 
+B{Q_{\nu-1/2}^{-\gamma-3/2}(\cosh (r/\ell_{eff}))
\over{\sinh^{3/2}(r/\ell_{eff})}}, \\ 
\psi=\psi_{\gamma}(t) &=&
 C{P_{1/2}^{\mu}(\tanh(Ht))\over{\cosh^{3/2}(Ht)}}
+D{P_{1/2}^{-\mu}(\tanh(Ht))\over{\cosh^{3/2}(Ht)}}, 
\end{eqnarray}
with
$\mu:=\sqrt{9/4+\lambda_{\gamma}^{2}}$, $\nu:=\sqrt{m^{2} l_{eff}^{2}+4}$ 
and $\lambda^2=\lambda_{\gamma}^{2}:=\gamma(\gamma+3)$.
The relative magnitude of the coefficients $A$ and $B$
are to be determined by the boundary conditions. 
The coefficients $C$ and $D$ are to be determined by the initial
condition. A natural initial condition in the scenario of creation of
the brane universe\cite{sasaki} would be to require $\dot\psi_\gamma=0$
at $t=0$.
However, here we leave the initial condition unspecified.


First, we consider the eigen-function $u_\gamma(r)$.
The boundary conditions to be satisfied are
\begin{eqnarray}
\partial_r u_{\gamma}|_{r=r_{0}}&=&0 \quad (\mbox{$Z_{2}$-symmetry}), 
\\
u_\gamma|_{r=0}&=&0 \quad (\mbox{Regularity at the origin}) .
\end{eqnarray}
Because $Q_\alpha^\beta(z)$ is singular as $z\to1$,
the regularity condition at the origin $r=0$ implies $B=0$.
In addition, $P_{\alpha}^{\beta}(z)$ behaves as $(z-1)^{-\beta/2}$ in
the limit $z\to1$. Hence we must have $\gamma>0$, i.e.,
$\lambda_\gamma^2>0$. 
On the other hand, from the asymptotic form of $P_\alpha^\beta(z)$
for $z \gg 1$, corresponding to $r \rightarrow \infty$,
we find
\begin{equation}
{P_{\nu -1/2}^{-\gamma-3/2}(\cosh(r/\ell_{eff}))
\over{\sinh^{3/2}(r/\ell_{eff})}}
\rightarrow z^{-\nu-2}+ z^{\nu-2},
\end{equation}
where $z=\cosh(r/\ell_{eff})$.
For $|\nu|<2$ ($m^{2}<0$) which we assume, the solution is damped as
$r\rightarrow \infty$. Thus it is possible for the solution to have at
least one extremum where $\partial_r u_\gamma=0$.
Therefore, the eigen-function $u_\gamma$ is given by
\begin{equation}
u_{\gamma}(r)=
{P_{\nu-1/2}^{-\gamma-3/2}(\cosh(r/\ell_{eff}))\over
\sinh^{3/2}(r/\ell_{eff})}\,.
\label{ugammasol}
\end{equation}
Given the location of the brane, the eigen-value $\gamma$ is determined
by the boundary condition at the brane, 
\begin{equation}
(\nu+2)z_{0}P_{\nu-1/2}^{-\gamma-3/2}(z_0)
=(\nu+\gamma+2)P_{\nu+1/2}^{-\gamma-3/2}(z_0)\,.
\label{location}
\end{equation}
Some examples of the behavior of $u_\gamma(r)$ are
shown in Fig.~1.
In extreme cases of the model parameters, the above equation can
be solved analytically. We will come back to this issue shortly.

\begin{minipage}{75mm}
\centerline{\epsfxsize=75mm\epsfbox{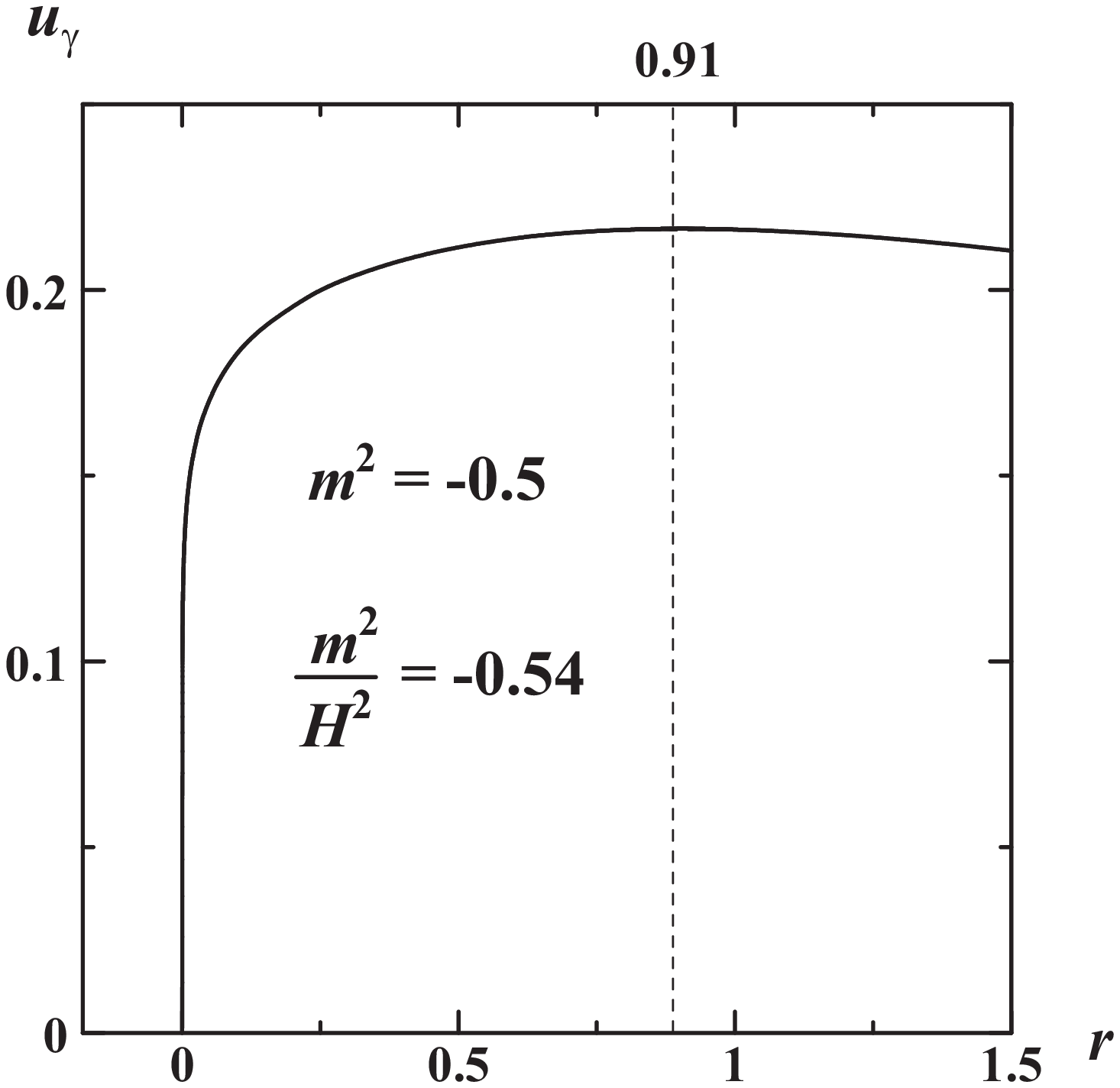}}
\end{minipage}\hspace{5mm}
\begin{minipage}{75mm}
\centerline{\epsfxsize=73mm\epsfbox{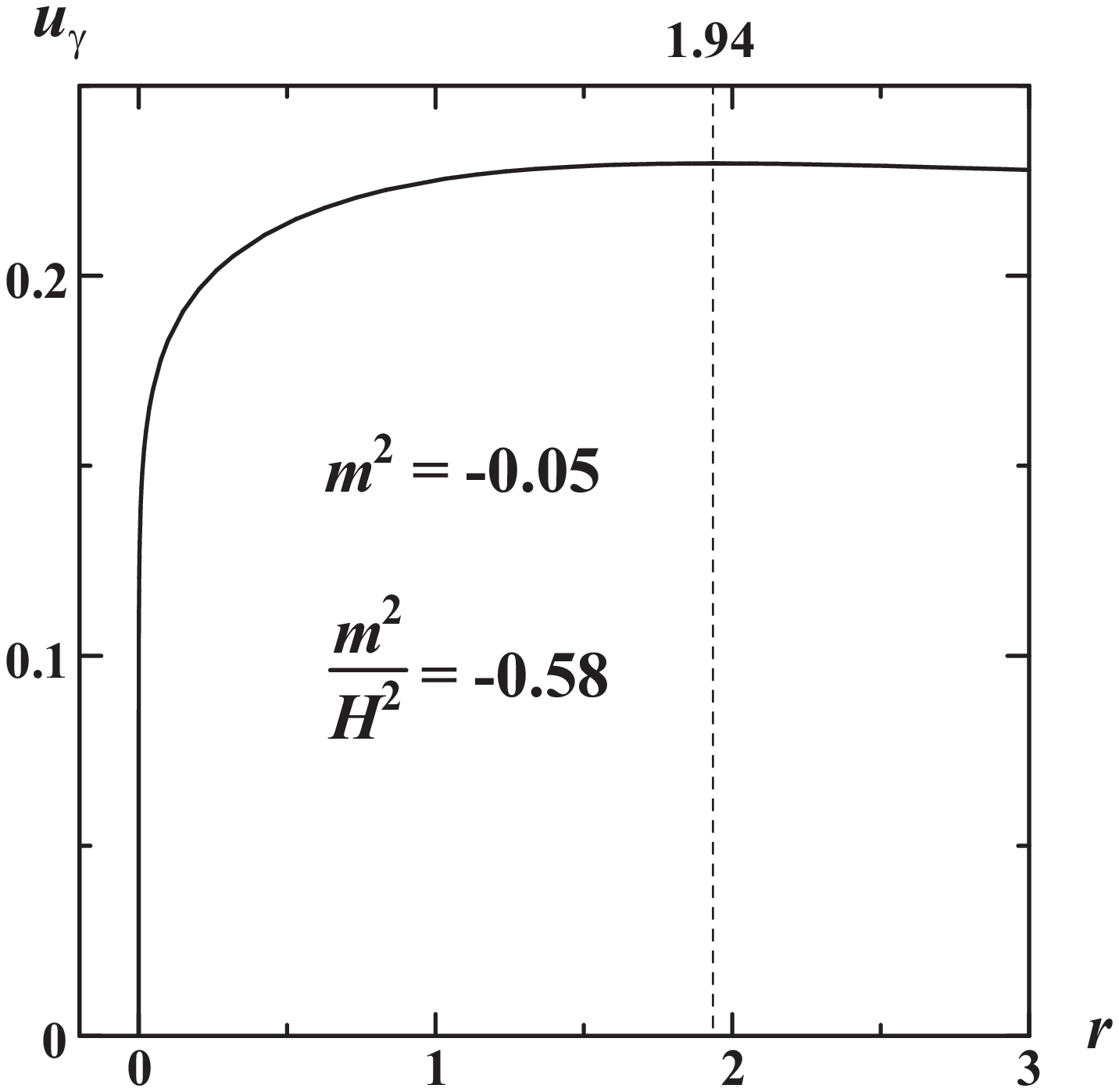}}
\end{minipage}

\vspace{5mm}
  \hspace*{10mm}
  \begin{list}{}{}
  \item[\hspace*{-5mm}Fig.~1~] Radial functions for a couple of model
     parameters in units of $\ell_{eff}=1$. The vertical scale is
     arbitrary.\\ The dotted vertical line indicates the location of
     the brane. In both cases, the eigen-value is $\gamma=0.1$.
  \end{list}
\vspace{5mm}

Next, we consider the time-function $\psi_\gamma(t)$.
For $Ht \gg 1$, we have
\begin{equation}
{P_{1/2}^{\pm\mu}(\tanh(Ht))\over\cosh^{3/2}(Ht)} \propto
\exp\left[\left(\pm\mu-{3\over2}\right)Ht\right],
\end{equation}
Noting that $\mu=\sqrt{9/4+\lambda_\gamma^2}$
and the condition $\lambda_\gamma^2>0$,
we find the solution approaches asymptotically to
\begin{eqnarray}
  \psi_\gamma(t)\to C{P_{1/2}^{\mu}(\tanh(Ht))\over{\cosh^{3/2}(Ht)}}
\propto \exp\left[\left(\mu-{3\over2}\right)Ht\right].
\end{eqnarray}
Thus slow-rolling occurs if the eigen-value $\lambda_\gamma^2$
satisfies the condition
\begin{equation}
0<\lambda_{\gamma}^{2} \ll 1 \Longleftrightarrow 0<\gamma \ll 1.
\label{slowrollcond}
\end{equation}
We have found numerically that there exists a solution that satisfies
the above condition for several examples of the model parameters
satisfying $|m^2|/H^2\ll 1$ (see Fig.~1 for a couple of examples).
To show analytically the existence of such a solution,
let us consider a couple of cases with extreme values of the
model parameters.

First, we consider the case when the vacuum energy $V_0$ almost
cancels out the cosmological constant $\Lambda_5$.
Namely,
\begin{equation}
\left|{\Lambda_{5,eff}\over\Lambda_5}\right|
={|\Lambda_{5}+\kappa_5^{2}V_0| \over |\Lambda_5|} 
=z_0^2-1\ll 1.
\end{equation}
Then the effective curvature radius $\ell_{eff}$ 
becomes very large compared with $\ell$.
In this case, Eq.~(\ref{location}) gives
\begin{equation}
z_{0}=z_{0,s}+O((z_{0}-1)^{2}),
\end{equation}
where
\begin{equation}
z_{0,s}=1+{4\gamma(5+2\gamma)\over{32+13\gamma-4(2+\gamma)\nu^{2}}}.
\end{equation}
Therefore, using Eq.~(\ref{srcondition}), 
we find
\begin{equation}
{|m^{2}|\over H^{2}}=5\gamma
+{(49+4\nu^{2}) \gamma^{2}\over{32-8\nu^{2}}}
+O(\gamma^{3}).
\end{equation}
Thus in the leading order we obtain
\begin{eqnarray}
  \gamma={|m^2|\over 5H^2}\,.
\end{eqnarray}
Hence, provided $|m^2|/H^2\ll1$, the slow-roll
condition (\ref{slowrollcond}) is satisfied.

Next, we consider the opposite case when $V_0$ is very small;
 $\kappa_5^{2}V_0/|\Lambda_5|\ll1$.
In this case, $\ell_{eff}$ is approximately equal to $\ell$.
The condition $|m^2|/H^2\ll 1$ implies
 $z_0=\cosh(r_0/\ell_{eff})\gg1$ and $|m^2|\ell_{eff}^2\ll |m^2|/H^2$.
 Then Eq.~(\ref{location}) is solved to give
\begin{eqnarray}
z_{0}^2={(\nu+\gamma+1)(\nu^{2}+(\nu-4)\gamma-4)
\over{4(\nu^{2}-3\nu+2)}}\,,
\end{eqnarray}
where $\nu=\sqrt{4+m^2\ell_{eff}^2}\approx2+m^2\ell_{eff}^2/4$.
Therefore, in the leading order we obtain
\begin{equation}
\gamma={|m^2|\over6H^2}-{|m^2|\ell_{eff}^2\over3}\,.
\end{equation}
Thus the slow-roll condition (\ref{slowrollcond})
is satisfied also in this case if $|m^2|/H^2\ll1$.

Although we have no analytical proof, the above results, together
with some numerical examples we have checked, strongly indicate
that the slow-roll solution exists for all possible values
of $\kappa_5^2V_0$ provided $|m^2|/H^2\ll 1$.

Finally, for completeness, we evaluate the 
$\partial_{\chi}^{2}\phi$ term in $E_{tt}$ given by Eq.~(\ref{weyl}). 
Noting that $\chi=r-r_0$ in the present model, we calculate 
$H^{-2}\partial_r^{2}\phi/\phi$ on the brane. 
Using Eq.~(\ref{location}) and the following recursion formula, 
\begin{equation}
(\alpha-\beta+1)P_{\alpha+1}^{\beta}(z)
-(2\alpha+1)z P_{\alpha}^{\beta}(z)
+(\alpha+\beta)P_{\alpha-1}^{\beta}(z)=0,
\end{equation}
it is expressed exactly as
\begin{equation}
{\partial_r^{2}\phi\over{H^{2}\phi}}=
\lambda_{\gamma}^{2}-{|m^{2}|\over H^{2}}\,.
\end{equation}
Hence $E_{tt}$ can be consistently neglected in 
the slow-roll situation when $|m^2|/H^2\ll 1$.

\subsection{Uniqueness of the solution}

We have found there exists at least one regular solution 
that satisfies the adequate boundary condition which
gives rise to slow-roll inflation on the brane, provided
the model parameters are chosen such that $|m^2|/H^2\ll 1$.
We now ask if our solution is unique or not.
If not, and if another solution happens to violate the slow-roll
condition, inflation on the brane would not stably last.

To see if there is such a possibility, we first rewrite the
radial eigen-value equation (\ref{req}) in the
standard Schr\"odinger form. 
To do so, we introduce the conformal radial coordinate $\eta$ 
through $dr/R(r)=d\eta$, 
where $R(r)=\ell_{eff}\sinh(r/\ell_{eff})$.
Then the metric (\ref{metric}) is expressed as
\begin{equation}
ds^{2}=R^{2}
\left(d\eta^{2}-H^{2}dt^{2}+\cosh^{2}(Ht)\,d\Omega_{(3)}^{2}\right),
\end{equation}
where
\begin{equation}
R(\eta)={\ell_{eff}\over{\sinh(|\eta|+\eta_0)}} \quad
 (-\infty<\eta<+\infty),
\end{equation}
and $\eta_0$ is defined by the equation
\begin{equation}
\sinh(\eta_0)={1\over \sinh(r_0/\ell_{eff})}=H\ell_{eff}.
\end{equation}
Then putting
$u_\gamma=R^{-3/2} f(\eta)$,
Eq.~(\ref{req}) becomes
\begin{equation}
-f''+\tilde{V}f=-\lambda_\gamma^{2}f\,,
\end{equation}
where the prime denotes the $\eta$-derivative and
\begin{eqnarray}
\tilde{V}&=&{(R^{3/2})''\over{R^{3/2}}}+m^{2}R^{2} 
\nonumber\\
&=& {9\over4}+{{15+4m^{2}\ell_{eff}^{2}}\over 4 \sinh^{2}(|\eta|+\eta_0)}
-3\coth(|\eta|+\eta_0) \delta(\eta).
\end{eqnarray}
For $m^2\ell_{eff}^2>-4$, this potential is of the volcano-type. 
Hence the solution
we have found corresponds to the unique bound state solution
with $E=-\lambda_\gamma^2<0$. Thus the solution is unique 
when $m^2\ell_{eff}^2>-4$.

On the other hand, when $m^2\ell_{eff}^2<-4$,
our solution might not be unique, depending on the location of
the brane. To investigate this case in more detail,
it is useful to analyze the behavior of the function $u_\gamma(r)$ 
given by Eq.~(\ref{ugammasol})
by artificially putting $\gamma=0$ (i.e., $\lambda_\gamma^2=0$). 
The function $u_0(r)$ can be expressed in terms of the
elementary functions as
\begin{eqnarray}
u_0(r)={\sqrt{2/\pi}\over |\nu|(|\nu|^{2}+1)}\,
{\sin(|\nu| \, r/\ell_{eff})
-|\nu|\cos(|\nu|\,r/\ell_{eff}) \tanh(r/\ell_{eff})
\over\sinh^{2}(r/\ell_{eff})\tanh(r/\ell_{eff})} \,,
\end{eqnarray}
where $|\nu|=\sqrt{|m^2|\ell_{eff}^2-4}$.
We then immediately see that the first node appears
in the region $\pi<|\nu|r/\ell_{eff}<3\pi/2$.  
Thus, if the location of brane $r_0$ satisfies
$r_0/\ell_{eff} <\pi/|\nu|$, $u_0$ has no node.
For a large value of $|\nu|$, Eq.~(\ref{srcondition})
with the requirement $|m^2|/H^2\ll 1$ gives
\begin{eqnarray}
\left({r_0\over\ell_{eff}}\right)^2(|\nu|^2+4)={|m^2|\over H^2}\,.
\end{eqnarray}
Thus the above nodeless condition is safely satisfied.
Since $u_0(r)$ corresponds to the zero energy solution,
this implies there is no bound state solution
in the energy range $0>E>-\lambda_\gamma^2$.
Hence our solution turns out to be unique also in this case.

\section{Summary and discussion}
We have investigated the possiblity of brane-world inflation 
driven solely a bulk gravitational scalar field. 

First we have clarified general (sufficient) conditions for inflation to
occur on the brane by analyzing the effective 4-dimensional Einstein
equations. Namely, we have found that the standard slow-roll inflation
can occur if the bulk scalar field is sufficiently slowly varying
both in space and time near the brane. 

Then, we have modeled the effective potential of the gravitational
scalar field by an new inflation type potential, and
looked for a spherically symmetric 5-dimensional solution 
perturbatively. In our model, the lowest order solution is given by
the AdS bulk with the de Sitter brane as the boundary,
and non-trivial behavior of the scalar field appears at the first order.
We have found there exists a regular solution for the scalar field 
in the separable form with respect to $r$ and $t$, where $r$ is the
5th dimensional radial coordinate.
The solution we have obtained gives slow-roll inflation on the brane
and is found to be unique, provided that $|m^{2}|/H^{2}\ll 1$,
which is the same condition for slow-roll inflation 
as in the usual 4-dimensional theory.

Although our solution is valid only perturbatively, we expect
a regular spherically symmetric solution to exist as long as
the scalar field potential is sufficiently flat.
Then, an immediate question is whether our model can give
the standard inflationary quantum fluctuation spectrum.
In this respect, very recently, Kobayashi, Koyama, and Soda
have calculated quantum fluctuations of a 5-dimensional
minimally coupled massless scalar field, and concluded
that the Kaluza-Klein modes give negligible contributions
to the quantum fluctuation spectrum induced on the 
brane\cite{shinpei}.
Hence we expect the same conclusion holds also in the present
model.

However, the situation seems to change drastically at the end
of inflation when the scalar field undergoes damped
oscillations. As one can notice from the discussions given in
Sec.~\ref{solution}, the radial function can be regular at the
origin only if $m^2<0$. Thus there exists no regular
solution in the separable form if $m^2>0$. This implies
that any regular solution must be in the form,
$\sum_\gamma \psi_\gamma(t)u_\gamma(r)$, where the sum extends
over infinite numbers of $\gamma$. Then there seems no apparent
reason why the sum should extend only over functions
of $t$ and $r$. In other words, it may be that an infinitesimal
non-spherical perturbation (the so-called Kaluza-Klein mode) 
can grow indefinitely.
This may cause a problem in our scenario. But at the same time,
if the non-spherical perturbations saturate at certain level,
this rather chaotic behavior may give rise to an efficient mechanism
of reheating. Furthermore, the Kaluza-Klein excitations left
in the subsequent universe may become a good candidate for the
cold dark matter of the universe.
Certainly, this issue deserves further studies.

Another remaining issue is the naturalness of the model parameters.
If we assume $\kappa_5^2V_0$ is of the same order of $\Lambda_5$,
we have $\ell_{eff}\gtrsim\ell$. Since the location of the brane
$r_0$ is expected to be greater than $\ell$ by factor of a few at
least, in order for our classical
picture of the infinitely thin brane to be valid, $r_0$ cannot be
smaller than $\ell_{eff}$. Hence the condition $|m^2|/H^2<1$
implies $|m^2|\ell_{eff}^2\ll1$ (see Eq.~(\ref{srcondition})). That is,
the mass parameter in our theory must be fine-tuned to a very
small value compared with the natural mass scale of the theory.
This is the same problem one encounters in 4-dimensional models
of inflation. However, if we recall that our model corresponds
to a conformally transformed scalar-tensor theory,
this problem may be solved in the original conformal frame.
Investigations in this direction is also left for future work.
    
\acknowledgements
{We would like to thank U. Gen, S. Kobayashi, N. Sago, J. Soda, 
T. Tanaka, and J. Yokoyama for discussions. Especially, we are 
grateful to T. Tanaka for very useful suggestions and comments
at early stages of this work.
This work was supported by the Research Funds of
the Yamada Science Foundation.}

\end{document}